\documentclass[paper,notoc]{JHEP3}
\usepackage{epsfig,cite}
\usepackage{amsbsy}
\usepackage{amsmath}
\usepackage{epsfig}
\usepackage{graphicx}

\title{\boldmath
The NNLO gluon fusion Higgs production cross-section with many heavy quarks.  
}

\author{Charalampos Anastasiou\\
  Institute for Theoretical Physics, ETH Zurich,\\
  8093 Zurich, Switzerland\\
  E-mail: \email{babis@phys.ethz.ch}}
\author{Radja Boughezal\\
  Institute for Theoretical Physics,  University of Zurich \\
Winterthurerstr. 190, 8057 Zurich, Switzerland\\
  E-mail: \email{radja@physik.uzh.ch}}
\author{Elisabetta Furlan\\
  Institute for Theoretical Physics, ETH Zurich,\\
  8093 Zurich, Switzerland\\
  E-mail: \email{efurlan@phys.ethz.ch}}

\abstract{
We  consider  extensions of the Standard Model with a number of
additional heavy quarks 
which couple to the Higgs  boson via top-like Yukawa interactions. 
 We  construct an effective theory valid for a Higgs  boson mass  which  is  lighter than 
twice the lightest heavy quark mass and  compute the corresponding Wilson coefficient through NNLO. 
We  present numerical results for the gluon fusion cross-section at the  Tevatron 
for  an extension of the Standard Model with a  fourth generation of
heavy quarks. 
The  gluon fusion cross-section is enhanced by a  factor of roughly 9
with respect to the Standard Model value.  Tevatron experimental data  
can place stringent  exclusion limits for the Higgs mass in this model.
}
\preprint{{ }}

\begin{document}

\section{Introduction}
\label{sec:intro}
The discovery of the Higgs boson will introduce a new era in particle  physics.  
The long standing theoretical problem of 
understanding  the mechanism for electroweak symmetry breaking  
will be  tackled for the first time using direct 
experimental findings. The measurement of the Higgs boson mass  
and the production cross-sections of its various signatures 
will be important  constraints in formulating 
a theory of particle interactions at high energies.

The interaction of the Higgs boson and  gluons is particularly important  at  hadron 
collider experiments. In the Standard Model (SM), the gluon fusion 
cross-section is the  largest  among  all  production cross-sections. 
The LHC will be  able  to discover the SM Higgs boson in this production channel for the full 
range of its allowed  mass values.   
The branching ratios for the decays of the SM Higgs boson are dominated  by other  Higgs boson interactions 
which involve  the bottom quark and electroweak gauge bosons.  However, 
the Higgs-gluon interaction is still not negligible; a fraction of up to about $7\%$ of Higgs 
bosons may decay to gluons, depending on the Higgs boson mass. 

Given that gluons are massless, a Higgs-gluon interaction arises as a loop effect via 
other massive coloured particles which couple to the Higgs boson. 
Physics beyond  the Standard Model can alter significantly the  strength of this interaction in various 
ways.  One possibility is that new coloured particles are not much heavier than the 
top quark, and their contribution is therefore not suppressed. 
A second possibility is that new coloured particles may  be heavier than the top-quark but they have 
an enhanced Yukawa coupling to the Higgs. A third possibility is that such particles are quite 
heavier than the top-quark, but their multiplicity is  large, thus building up a  significant cumulative contribution. 

A Higgs  boson is often assumed  to be lighter than about twice the mass of the  top-quark 
and twice the  mass of new undiscovered  particles which are hypothesized  in extensions of 
the Standard Model.  Light new particles are  hard to accommodate given the vigorous experimental 
searches  for new  physics at LEP and the Tevatron.  On the other hand, they 
cannot  be very heavy or, alternatively, they must  have  a rather strong  Higgs coupling 
if they contribute in reducing the fine tuning of the Higgs mass. Therefore, it  is important  
to calculate their  contribution in the gluon fusion process as  well as in the decay of a Higgs boson to 
gluons.  

Assuming a Higgs  boson which is lighter than production thresholds of new heavy particles, we  can factorize  
the effect of new physics and QCD  in the process $gg \to H$, by means of  an effective field  
theory where the top-quark and all other possible heavy coloured 
states which couple  to the Higgs boson are integrated out. 
The effect  of these  heavy particles is included  in the Wilson coefficients of an
effective theory with operators of the  Higgs boson and  light quarks  and gluons. 

The possibilities for viable extensions  of  the Standard Model which alter  the Higgs-gluon interaction 
are many, and an equal number of  matching  calculations  is required  for their study. 
This is a  rather easy task at leading order in the  strong  coupling. However, experience  from the 
Standard Model shows that  a precise  estimate  of the gluon-fusion cross-section and  
the Higgs decay width to gluons requires  calculations  through next-to-next-to-leading-order (NNLO) in the 
strong  coupling.

In this paper, we consider extensions of the Standard Model with additional heavy quarks. 
We  assume that these quarks  have a Higgs Yukawa 
interaction of the same type  as Standard Model quarks.  The existence of such quarks has dramatic implications for the 
Higgs production  cross-section  in gluon fusion. At leading order, and in the limit  where the heavy quarks are much 
heavier than half the mass of the Higgs  boson, the cross-section scales  as  $n_h^2$, where $n_h$ is the number of heavy quarks. Current 
measurements at the Tevatron~\cite{Aaltonen:2010cm} and early data from the LHC can therefore  constrain severely such models.  
We first construct an effective Lagrangian integrating out the top-quark and the additional heavy quarks.  We compute the Wilson 
coefficient  of the Higgs-gluon effective  interaction through NNLO in the strong  coupling  expansion. Finally, we present numerical results for the 
gluon fusion cross-section at the Tevatron in a specific model with a fourth quark generation.

\section{The effective Lagrangian}

We consider an arbitrary extension of the SM through new heavy 
quarks transforming under the fundamental representation of the QCD 
gauge group $SU(3)
%% _{QCD}
$. The number of heavy quarks, including 
the top, is $n_h$. We will denote their mass by $m_q$, with 
$\; q=1 \ldots n_h$. 
We assume that the new quarks, as the SM top, couple to the Higgs 
boson $H$ through their mass.
Therefore, the Lagrangian we begin with is
\begin{equation}
\label{eq:lag_full}
  {\cal L}
  =
      {\cal L}^{n_l}_{QCD}
  + \sum_{q=1}^{n_h}  \bar{\psi}_q 
  	  \left(  
  			i D \!\!\!\!\slash - m_q 
  	   \right) 
       \psi_q
    + {\cal L}_Y
    \quad, \quad
    {\cal L}_Y
    =
    - \frac{ H }{ v }
    	\sum_{q=1}^{n_h}  
    				m_q  \bar{\psi}_q  \psi_q
       \; .
\end{equation}
Here $D_{\mu}$ is the covariant derivative in the fundamental representation 
and $ {\cal L}^{n_l}_{QCD}$ is the QCD Lagrangian with only the 
$n_l$ flavours of light quarks. We take these quarks to be massless. 

We focus on the changes that the heavy quarks induce on the Higgs production 
through gluon fusion.
When the quarks that couple to the Higgs boson are heavier than half the 
Higgs boson mass, we can integrate them out.
In this limit, we can replace the original Lagrangian (\ref{eq:lag_full}) 
with an effective Lagrangian
\begin{equation}
	{\cal L}^{eff} 
	=
	   {\cal L}_{QCD}^{eff,n_l}
     - C_1 \,\frac{H}{v} \,{\cal O}_1
     \;.
\label{eq:Leff}
\end{equation}
$C_1$ is the Wilson coefficient~\cite{Wilson:1969zs} relative to the 
only dimension-four local operator 
${\cal O}_1$ that arises when we integrate out the heavy quarks
and all the quarks remaining are massless~\cite{Spiridonov:1984br}, 
\begin{equation}
	{\cal O}_1 
	= 
	\frac{1}{4} \,{G'}^a_{\mu\nu} {G'}^{a\mu\nu} 
	\; .
\end{equation}
In this expression, ${G'}^a_{\mu\nu}$ is the field strength tensor in the effective theory. 
In Eq.~(\ref{eq:Leff}), 
$  {\cal L}_{QCD}^{eff,n_l}$ describes the interactions among light 
quarks. It has the same form as
$  {\cal L}_{QCD}^{n_l}$, but with different parameters and field normalizations 
because of the contributions from heavy quarks loops.  
We relate the parameters in the effective theory to the parameters in 
the full theory through multiplicative decoupling constants $\zeta_i$. 
We will denote quantities in the effective theory with a prime.
The derivation of the decoupling constants is reviewed in \cite{Steinhauser:2002rq}. In Section \ref{renormalization}, 
we describe the main steps of their calculation and give the relevant results.

\section{Method}
\label{sec:comp1}
We compute the Wilson coefficient $C_1$ up to three
loops. Diagrams containing both the heavy mass
scales appear for the first time at the three-loop order. 
We start from the bare amplitude
${\cal M }^0_{gg \to H}$ 
for the process $g g \to H$ in the full theory, 
\begin{equation}
	{\cal M }^0_{gg \to H} 
	\equiv 
	{\cal M}^{0,a_1 a_2}_{\mu_1\mu_2}(p_1, p_2) 
		\epsilon^{\mu_1}_{a_1} \epsilon^{\mu_2}_{a_2}
	\;.
\end{equation}
Here, $p_1$ and $p_2$ are the momenta of the two gluons 
with polarizations $\epsilon^{\mu_1}_{a_1}$ and  
$\epsilon^{\mu_2}_{a_2}$.
This amplitude is related to the bare Wilson coefficient $C^0_1$ by~\cite{Steinhauser:2002rq}
\begin{equation}
	\frac{\zeta^0_3 C^0_1}{v}
 	=
	\frac{\delta^{a_1 a_2}
			\left( g^{\mu_1\mu_2} (p_1\cdot p_2) - p_1^{\mu_2} p_2^{\mu_1} \right)
			}
			{ (N^2-1) (d-2) {(p_1\cdot p_2)^2} } 
	{\cal M}^{0,a_1 a_2}_{\mu_1\mu_2}(p_1, p_2)
   \bigr\rvert_{p_1 = p_2 = 0}
   \;.
\label{eq:C01}
\end{equation}
$N$ is the number of colours and $d=4-2\epsilon$ is the dimension of
space-time. Bare quantities are denoted by the superscript $``0"$. 
The factor $\zeta^0_3$ is the bare decoupling
coefficient by which the bare gluon field 
${G'}^{0,a}_{\mu} $ 
is rescaled in the effective theory,
\begin{equation}
	{G'}^{0,a}_{\mu} 
	= 
	\sqrt{\zeta^0_3} \,{G}^{0,a}_{\mu} 
	\; .
\end{equation}

We  generate  the Feynman diagrams  ${\cal F}$ 
for the amplitude through three loops using QGRAF~\cite{Nogueira:1991ex}. 
We then perform an expansion of all diagrams in the external momenta $p_1,p_2$, 
by applying the following  differential operator~\cite{Fleischer:1994ef} 
to their integrand: 
\begin{equation}
	{\cal D} {\cal F} 
	= 
	\sum_{n=0}^\infty (p_1 \cdot p_2)^n \left[ {\cal D}_n {\cal F} \right]_{p_1 = p_2 =0}, 
\end{equation}
with 
\begin{equation}
{\cal D}_0 =1, 
\quad 
{\cal D}_1 =\frac{1}{d} \Box_{12},
\qquad
 {\cal D}_2 =-\frac{1}{2(d-1)d(d+2)} \left\{ 
\Box_{11} \Box_{22} - d\; \Box_{12}^2
\right\}, 
\end{equation}
and $ \Box_{ij} \equiv g^{\mu \nu}\frac{\partial^2}{\partial p_i^\mu \partial p_j^\nu}$. \\
Differential operators of higher orders are not needed for the expansion 
in the external momenta at leading order.

After Taylor expansion, all the Feynman diagrams are expressed in terms 
of one-, two- and three-loop vacuum bubbles by using linear transformations 
on the loop-momenta $k_i$: 
\begin{eqnarray}
	I_{1}\left[ \nu_1 \right] 
	& \equiv & 
	\int \!\! \frac{d^dk_1}{i \pi^{d/2}}
				 \frac{1}{{\cal P}_1^{\nu_1}} 
	\;, \\
%\end{equation} %two-loop, %\begin{equation}
	I_{2}\left[ \nu_1, \nu_2, \nu_5 \right] 
	& \equiv &
	\int \!\! \frac{ d^d k_1 d^d k_2} {(i \pi^{d/2})^2}
			 \frac{1}{  {\cal P}_1^{\nu_1}  {\cal P}_2^{\nu_2}  {\cal P}_5^{\nu_5}} 
	\;, \\
%\end{equation}%and  three-loop %\begin{equation}
	I_{3a}\left[ \nu_1, \nu_2, \nu_3, \nu_5, \nu_6, \nu_7  \right] 
	& \equiv & 
	\int \!\! \frac{d^d k_1 d^d k_2 d^dk_3}{(i \pi^{d/2})^3}
		\frac{  1  }{  {\cal P}_1^{\nu_1}  {\cal P}_2^{\nu_2}  {\cal P}_3^{\nu_3} {\cal P}_5^{\nu_5}  {\cal P}_6^{\nu_6}  {\cal P}_7^{\nu_7}}
		\;, \\ 
%\end{equation}%\begin{equation}
	I_{3b}\left[ \nu_1, \nu_2, \nu_3, \nu_4, \nu_5, \nu_6  \right] 
	& \equiv &
	\int \!\! \frac{d^d k_1 d^d k_2 d^dk_3}{(i \pi^{d/2})^3}
	\frac{  1  }{  {\cal P}_1^{\nu_1}  {\cal P}_2^{\nu_2}  {\cal P}_3^{\nu_3} {\cal P}_4^{\nu_4}  {\cal P}_5^{\nu_5}  {\cal P}_6^{\nu_6}}
	\;, \\ 
	I_{3c}\left[ \nu_1, \tilde{\nu}_2, \tilde{\nu}_3, \nu_4, \nu_5, \nu_6  \right] 
	& \equiv &
	\int \!\! \frac{d^d k_1 d^d k_2 d^dk_3}{(i \pi^{d/2})^3}
		\frac{  1  }{  {\cal P}_1^{\nu_1}  \tilde{{\cal P}}_2^{\tilde{\nu}_2}  \tilde{{\cal P}}_3^{\tilde{\nu}_3} {\cal P}_4^{\nu_4}  {\cal P}_5^{\nu_5}  {\cal P}_6^{\nu_6}}
	\; ,
\end{eqnarray}
with  
\begin{equation}
\begin{array}{ccll}
{\cal P}_{1} &=& k_1^2- m_q^2 \;,  & \\
{\cal P}_{2} &=& k_2^2- m_q^2 \;,  & \tilde{{\cal P}}_{2} = k_2^2- m_{q'}^2 \;, \\
{\cal P}_{3} &=& k_3^2- m_q^2 \;,  & \tilde{{\cal P}}_{3} = k_3^2- m_{q'}^2 \;, \\
{\cal P}_{4} &=& (k_1 -k_2 + k_3)^2 - m_q^2 \;, \quad & \\
{\cal P}_{5} &=& (k_1 -k_2)^2 \;, & \\ 
{\cal P}_{6} &=& (k_2 - k_3)^2 \;,  & \\
{\cal P}_{7} &=& (k_3 - k_1)^2 \;,  & 
\end{array}
\end{equation}
and $\nu_i$, $\tilde{\nu}_i$ positive or negative integers. 
The third three-loop vacuum bubble $I_{3c}$ contains two heavy quarks of different 
mass, $m_q$ and $m_{q'}$.

We  perform a  reduction of the above integral topologies to master integrals 
using the algorithm of Laporta~\cite{Laporta:2001dd} and the
program AIR~\cite{Anastasiou:2004vj}. 
We find five master integrals, 
\begin{eqnarray}
 {\cal I}_1 &=& I_1[1] \nonumber \\ 
 				&=& - \left( m_q^2\right)^{1-\epsilon} \Gamma(-1+\epsilon) \; , \\
 {\cal I}_2 &=& I_{3a}[1, 0, 1, 1, 1, 0] \nonumber \\ 
 				&=& \left( m_q^2 \right)^{2-3\epsilon} 
 						\frac{ \Gamma^2(1-\epsilon) \Gamma(\epsilon) \Gamma^2(-1+2\epsilon) \Gamma(-2+3\epsilon)}
 								{\Gamma(2-\epsilon)\Gamma(-2+4\epsilon)} \; , \\
 {\cal I}_3 &=& I_{3b}[1, 1, 1, 1, 0, 0] \;, \\  
 {\cal I}_4 &=& I_{3c}[1, 1, 1, 1, 0, 0] \;, \\  
 {\cal I}_5 &=& I_{3c}[2, 1, 1, 1, 0, 0] \;.
\end{eqnarray}
Single-scale master integrals appear in the calculation of the SM Wilson 
coefficient~\cite{Chetyrkin:1997iv, Chetyrkin:1997un} and can be computed 
with MATAD~\cite{Steinhauser:2000ry}. 
For the remaining two-scale master integrals we used result from~\cite{Bekavac:2009gz}. 
We checked all the master integrals independently through sector decomposition 
with the program of Ref.~\cite{Anastasiou:2007qb}.\\
After these steps, the RHS of Eq.~(\ref{eq:C01}) becomes
\begin{eqnarray}
	\label{eq:C10m1m2}
	\frac{\zeta^0_3 C^0_1}{v}
 	&=&
 	\sum_{q=1}^{n_h}
 	\left\{
 	\frac{1}{3}
 	\left( \frac{\alpha_s^0 S_{\epsilon}}{\pi} \right) 
 	\left[
 			-1
 			+\epsilon \left[
 							1 + 2 \log (m_q^0) 
 				\right]
 				\phantom{\frac{1}{2}}
 				\right.
 	\right.
 				\nonumber \\
 		&&
 				\left.
 				\phantom{-----}
 			- 2 \epsilon^2 \left[
 					   \log^2 (m_q^0)
 					+ \log (m_q^0)
 					+\frac{\pi^2}{24}
 			\right]
 			+ {\cal O}(\epsilon^3)
 	\right]
 	\nonumber \\
 	&&
 	+
 	\left( 
 			\frac{  \alpha_s^0 S_{\epsilon}  }{  \pi  } 
 	\right)^2
	\left[
		-\frac{ 1 }{ 4 }
		+ \epsilon \left[
				\log (m_q^0) + \frac{31}{36}
		\right]
		+ {\cal O}(\epsilon^2)
	\right]
	\nonumber \\
	&&
	+
	\left( 
			\frac{\alpha_s^0 S_{\epsilon}}{\pi} 
	\right)^3 
	\left[
			-\frac{1}{32 \epsilon^2}
			+\frac{1}{\epsilon} 
			   \left[
			   		\frac{3 \log (m_ q^0)}{16}-\frac{223}{576}
			   	\right]
			 \right.
			 \nonumber \\
			 && \left.
			 \phantom{------}
			+ n_l \left[
					-\frac{ 5 }{ 144 } \frac{1}{\epsilon}
					+\frac{103}{864} + \frac{5 \log (m_ q^0)}{24}
			  \right]
			  \right.
			 \nonumber \\
			 && \left.
		\left.
			 \phantom{------}
			-\frac{9}{16} \log^2(m_q^0)+\frac{223 \log (m_q^0)}{96}-\frac{\pi ^2}{128}+\frac{5975}{3456}
			+ {\cal O}(\epsilon)
	\right]
\right\} \nonumber \\
%% PIECE FROM THE MIXED DIAGRAMS
 & &
 -\left( \frac{\alpha_s^0 S_{\epsilon}}{\pi} \right)^{\!\!3} \!\!
 \sum_{q>q'}
 \left\{
 	\frac{1}{16 \epsilon^2}
 	-\frac{1}{16\epsilon}
 		\left[
 				{3 \left(\log (m_q^0)+\log (m_{q'}^0)\right)+\frac{89}{18}}
 		\right]
 	+\frac{1}{2} \log (m_q^0) \log (m_{q'}^0)
 	\right.\nonumber\\
 	&& \left.
 	+\frac{89}{96} \left[\log (m_q^0)+\log (m_{q'}^0) \right]
 	+\frac{5}{16} \left[\log ^2(m_q^0)+\log ^2(m_{q'}^0)\right]
 	+\frac{1051+27 \pi ^2}{1728}
 	+ {\cal O}(\epsilon)
 	\right\}
\;. \nonumber \\
\end{eqnarray}
The first sum in this expression runs over all single-scale diagrams and corresponds to $n_h$ copies 
of the SM Wilson coefficient. The second sum accounts for the 3-loop diagrams in which either 
of the two massive quarks couples to the Higgs boson. We already symmetrized it over $q, q'$. 
In Eq.~(\ref{eq:C10m1m2})
we introduced the factor 
\begin{equation}
	S_{\epsilon} 
 =
  e^{-\epsilon \gamma_E} 
  \left(  4 \pi  \right)^\epsilon  \,.
\end{equation}

\section{Decoupling and renormalization}
\label{renormalization}

The RHS of Eq.~(\ref{eq:C10m1m2})
contains the bare masses of the heavy 
quarks $m_q^0$ and the bare coupling constant
$\alpha_s^0$ in the full theory; 
$C_1^0 =C_1^0( \alpha_s^0, m_q^0) $. 
The bare strong coupling in the full theory is related to the 
bare strong coupling in the effective theory $\alpha_s^{'0}$
by the decoupling constants $\zeta_g^0$% and $\zeta_{m_q}^0$
~\cite{Steinhauser:2002rq, Chetyrkin:1997un}, 
\begin{equation}
	\label{eq:decouple_bare}
	\alpha_s^{'0}   =   (\zeta_g^0)^2 \alpha_s^{0} \quad . \quad
%	m_q^{'0}   =   \zeta_{m_q}^0 m_q^0 \;.
\end{equation}
Similiarly,
\begin{equation}
	\alpha'_s  =   (\zeta_g)^2 \alpha_s \quad . \quad
%	m'_q   =   \zeta_{m_q} m_q \;.
\end{equation}
Using these relations, we obtain the bare Wilson coefficient as a 
function of the bare parameters in the effective theory and of the bare mass of the 
heavy quarks in the full theory, 
$C_1^0 =C_1^0( \alpha_s^{'0}, m_q^{0}) $. 
The bare parameters are related to the renormalized ones through multiplicative 
renormalization constants $Z_i$ as
\begin{eqnarray}
	\label{eq:renormalization_effective}
	\alpha_s^{'0}   &=&  \mu^{2 \epsilon} Z_{\alpha}' \alpha_s'(\mu) \quad , \quad
	\phantom{m_l^{'0}   =   Z'_{m_l} m'_l(\mu) \;,} \\
	\label{eq:renormalization_full}
	\alpha_s^{0}   &=&  \mu^{2 \epsilon} Z_{\alpha} \alpha_s(\mu) \quad , \quad
	m_q^{0}   =   Z_{m_q} m_q(\mu) \;.
\end{eqnarray}
All the parameters in Eq.~({\ref{eq:renormalization_effective})
are in the effective theory 
and all the parameters in 
Eq.~({\ref{eq:renormalization_full}) are in the full theory. 
Finally, we renormalize the Wilson coefficient itself through a 
renormalization factor $Z_{11}$~\cite{Spiridonov:1984br,Spiridonov:1988md,Chetyrkin:1996ke},
\begin{equation}
 C_1 = \frac{ 1 }{ Z_{11} } C_1^0 \;.
\end{equation}.

%%%%%%%%%%%%%%%%%%%%%%%%%%%%%%%%%%%%%%%%%%%%%%%%%%%%%%%%
%																															%
\subsection{Details of the calculation}																%
%																															%
%%%%%%%%%%%%%%%%%%%%%%%%%%%%%%%%%%%%%%%%%%%%%%%%%%%%%%%%

A convenient way to compute the gluon field decoupling 
$\zeta^0_3$ is through the relation
\begin{equation}
	\zeta^0_3 = 1 + \Pi^0_G(p=0) 
	\; ,
\end{equation}
where 
$ \Pi^0_G $
is the transverse component of the gluon self-energy in the full 
theory. This quantity is computed at zero external momentum.
Since we work in dimensional regularization, only diagrams containing  
at least one massive quark contribute.
We find only one diagram per 
heavy flavour at one loop and  seven at two loops. 
We employ the same calculation techniques as in Section~\ref{sec:comp1}.
Our result reads 
\begin{eqnarray}
\label{eq:zeta03}
 \zeta^0_3
 &=&
 1
 +
 \sum_{q=1}^{n_h}
 \left\{
	\left( \frac{\alpha_s^0 S_{\epsilon}}{\pi} \right) 
	\left[
			   \frac{1}{6 \epsilon}
			- \frac{\log (m_q^0)}{3}
			+	\epsilon \frac{\pi^2 + 24 \log^2(m_q^0) }{ 72 }
	\right]
	\nonumber \right.\\
	&&
	+
	\left.
	\left(  \frac{  \alpha_s^0 S_{\epsilon}  }{  \pi  } \right)^2 
	\left[
			   \frac{3}{32 \epsilon^2}
			 -\frac{1+24 \log (m_q^0)}{64 \epsilon}
			+\frac{3}{4} \log^2(m_q^0)
			+\frac{1}{16}\log (m_q^0)
			+\frac{91}{1152}
			+\frac{\pi^2}{64} 
	\right]	
\right\} \;.
\nonumber \\
\end{eqnarray}
%%%%%%%%%%%%%%%%%%%%%%%%%%%%%%%%%%%%%%%%%%%%%%%%%%%%%%%%%%%%

The bare decoupling parameter of the strong coupling constant, 
$ \zeta^0_g $, 
can be computed as
\begin{equation}
\label{eq:zeta0g_relations}
	\zeta^0_g = \frac{  \tilde{\zeta}^0_1  }
								{  \tilde{\zeta}^0_3  \sqrt{ \zeta_3^0}  }
\;,
\end{equation}
where 
$ \tilde{\zeta}^0_1 $, $ \tilde{\zeta}^0_3 $ 
and $ \zeta_3^0 $ 
are the bare decoupling constants of the gluon-ghost vertex, 
of the ghost field and of the gluon field respectively.\\
The bare decoupling constant of the ghost field is calculated 
from the ghost self-energy in a similar way as 
$ \zeta_3^0 $. At one loop, there is no diagram contributing to the 
ghost decoupling. At two loops, there is only one diagram per heavy
flavour. We  find
\begin{eqnarray}
\label{eq:zeta03tilde}
	\tilde{\zeta}_3^0 
	&=& 
	1
	+ 
% no one-loop contribution
	 \left(
	 	\frac{  \alpha_s^0 S_{\epsilon}  }{  \pi  }
	 \right)^2
	\sum_{q=1}^{n_h} 
	\left[
		-  \frac{ 3 }{ 64 \epsilon^2 }
		+  \frac{ 1 }{ \epsilon }
		\left(
			  \frac{ 5 }{ 128 }
			+\frac{ 3}{ 16 } \log (m_q^0) 
		\right)
		-
		\frac{ 89 + 6 \pi^2 }{ 768 }
		\right. \nonumber \\
		&  &
		\phantom{spaaaaaaaaaaaaaaaaaaaaaaaaaace}
		\left. 
		-  \frac{ 5  }{ 32 } \log(m_q^0)
		-  \frac{ 3 }{ 8 } \log^2(m_q^0)
	\right]
	\;.
\end{eqnarray}
%%%%%%%
%%%%%%%
The decoupling of the gluon-ghost vertex $\tilde{\zeta}^0_1$ 
is given by 
\begin{equation}
	\tilde{\zeta}_1^0 
	=
	1
	+  \Gamma^{0}_{\bar{\eta} G \eta }(0,0)
	\;,
\end{equation}
where $\Gamma^{0}_{\bar{\eta} G \eta }(p,p')$ 
is extracted from the 1PI amputated gluon-ghost 
Green function and $p$ and $p'$ are the incoming 
four-momenta of $\bar{\eta}$ and $G$ respectively. 
Again, this term receives no contribution at one loop. 
At two loops, there are 5 non-massless diagrams for each
heavy flavour. Two of them vanish because of colour, and 
the other three add up to zero. Therefore
\begin{equation}
\label{eq:zeta10tilde}
	\tilde{\zeta}_1^0 
	=
	   1
	+ {\cal O}(\alpha_s^3) \;.
\end{equation}
Inserting Eqs.~(\ref{eq:zeta10tilde}, \ref{eq:zeta03tilde}, \ref{eq:zeta03}) 
into the relation~(\ref{eq:zeta0g_relations}) we find
\begin{eqnarray}
	\zeta_g^0
	&=&
	   1
	+ \left( \frac{\alpha_s^0 S_{\epsilon}}{\pi} \right) 
		\left[
			-\frac{n_h}{12 \epsilon}
			+\frac{L_q^0}{6}
			- \epsilon \left(  \frac{ L_{2,q}^0 }{ 6 }+ n_h \frac{ \pi ^2 }{144}    \right)
		\right]
	\nonumber \\
	& & \phantom{1}
	+ \left( \frac{\alpha_s^0 S_{\epsilon}}{\pi} \right)^2 
		\left\{
			  \frac{n_h^2}{96 \epsilon^2}
			- \frac{n_h}{ 24 \epsilon}
				\left[   L_q^0   +   \frac{3}{4}   \right]
			+\frac{L_q^0 }{8}	 +   \frac{(L_q^0)^2}{24}
			+\frac{n_h}{24}
				\left[   L_{2,q}^0 + \frac{  11  }{  6  }    \right] 
			+ n_h^2 \frac{\pi^2 }{576}
		\right\} \;.
		\nonumber \\
\end{eqnarray}
Here we introduced the notation
\begin{equation}
	L_q^0 = \sum_{q=1}^{n_h} \log(m_q^0)
	\quad , \quad
	L_{2,q}^0 = \sum_{q=1}^{n_h} \log^2(m_q^0)
	\;.
\end{equation}

%%%%%%%%%%%%%%%%%%%%%%%%%%%%%%%%%%%%%%%%%%%%%%%%%%%%
We now renormalize the mass of the heavy quarks in the full theory 
according to Eq.~(\ref{eq:renormalization_full}).
The mass renormalization constants in the full theory $Z_{m_q}$ are obtained 
from the one- and two-loop corrections to the quark propagator \cite{Tarrach:1980up}. 
%%%%%%% New piece
We review here the main steps of this calculation. 
Let us denote the sum of all the one-particle irreducible (1PI) insertions into the quark 
propagator as $-i \Sigma_0(p)$,
 \begin{equation}
  -i \Sigma_0(p) = -i \Sigma^{1L}_0(p) -i \Sigma^{2L}_0(p) + \ldots \;,
 \end{equation} 
 where $ -i \Sigma^{1L}_0(p)$ in the sum of all the one-loop 1PI diagrams in the bare theory 
and so on.
The full quark propagator then reads
\begin{equation}
\label{eq:quark_propag1}
	\frac{i}
			{	\slash \!\!\! p - m_q^0 - \Sigma_0(p)	}
	\;.
\end{equation}
We can split $\Sigma_0(p)$ as
\begin{equation}
\label{eq:split_sigma}
	\Sigma_0(p) 
	= 
	\Sigma_{10}(p^2) + (\slash \!\!\! p -m_q^0) \Sigma_{20} (p^2)
	\;;
\end{equation}
conversely, the quantities
$ \frac{1}{m_0}		\Sigma_{10}$,
$ 	\Sigma_{20}  $
are extracted from the bare self-energy 
$\Sigma_0$
as
\begin{equation}
	\frac{1}{m_0}	\Sigma_{10}
	=
	\frac{1}{4}	\textrm{Tr}
	\left(
			\frac{ 1 }{ m_0 } \Sigma_0
		+	\frac{\slash \!\!\! p }{ p^2 }	\Sigma_0
	\right)
	\quad , \quad
	\Sigma_{20}
	=
	\frac{1}{4 p^2 }	\textrm{Tr}
	\left(
		\slash \!\!\! p		\Sigma_0
	\right)
	\;.
\end{equation}
Combining Eqs.~(\ref{eq:quark_propag1}) and~(\ref{eq:split_sigma}) we obtain
\begin{eqnarray}
\label{eq:ZM_renorm}
	m_q^0
	& = &
	m_q(\mu)
	\left[
		  1
		- \frac{1}{m_q} \Sigma_{ 1 }^{ 1L }
		-	\left(
				  \frac{1}{m_q} \Sigma_{ 1 }^{ 2L }
				+\frac{1}{m_q}   \Sigma_{ 1 }^{ 1L }   \Sigma_{ 2 }^{ 1L }
		    \right)
	\right]
	\;.
\end{eqnarray}	
The RHS of Eq.~(\ref{eq:ZM_renorm}) is expressed in terms of renormalized quantities.
On the other hand, the one-loop result for $\Sigma_{10}^{1L}$ originally  depends on 
the bare gauge parameter and on the bare strong coupling and is divided itself by 
the bare mass. Therefore, one also needs the one-loop 
renormalization of these three quantities in order to compute $Z_{m_q}$ through two loops.\\ 
We compute the self-energy for arbitrary external momentum $p$,  
%Since the external momentum acts as infrared regulator, we can 
and set the mass of the heavy quarks to zero. 
In the $\overline{{\rm MS}}$ scheme we recover the result~\cite{Tarrach:1980up}
\footnote{Note that in our conventions $d = 4-2 \epsilon$, while in Ref.~\cite{Tarrach:1980up}
$d = 4+2 \epsilon$. This explains the sign difference in the $1/\epsilon$ terms.}
\begin{equation}
\label{eq:Zmq}
	Z_{m_q}
	=
	   1
	- \frac{ \alpha_s(\mu) }{ \pi  }
		\frac{ 1 }{ \epsilon }
	+\left(      \frac{ \alpha_s(\mu) }{ \pi }     \right)^2
		\left[
			\frac{ 1 }{ \epsilon^2 }
			\left(
					\frac{ 45 - 2 n_f }{ 24 } 
			\right)
			+
			\frac{ 1 }{ \epsilon }
			\left(
					- \frac{ 101 }{ 48 }	+	\frac{ 5 }{ 72 } n_f
			\right)
	\right]
 \;.
\end{equation}
This relation holds both in the full theory, where the number of active flavours
$n_f$ is $n_f = n_l + n_h$, and in the effective theory. In this case, 
$n_f = n_l$ and we need to replace the renormalized strong coupling in 
the full theory with the one of the effective theory. \\
%It was shown in \cite{Chetyrkin:1997un} that the mass decoupling constant $\zeta_m$ receives the 
%first correction at $ {\cal O}(\alpha_s^2(\mu))$. This enters the coefficient 
%on the RHS of Eq.~(\ref{eq:ren_dec_masses}) as a finite term at 
%$ {\cal O}(\alpha_s^2(\mu))$. Such term does not contribute to the 
%final result, as it enters the three-loop Wilson coefficient only starting 
%from ${\cal O}(\epsilon)$.\\
So far, the coefficient for the mass decoupling and renormalization still depends on the renormalized 
strong coupling in the \textit{full} theory. 
We decouple it using the relation~\cite{Steinhauser:2002rq}
\begin{equation}
	\alpha_s'(\mu)
	=
	\frac{   Z_{\alpha} (\zeta_g^0)^2    }
			{   Z'_{\alpha}   }
	\alpha_s(\mu)
	=
	 \zeta_g^2		\alpha_s(\mu)
	\;.
\end{equation}
The strong coupling renormalization constants are  
related to the coefficients of the $\beta$ function as
\begin{equation}
\label{eq:Zalpha}
	Z'_{\alpha}
	= 
	   1
	- \frac{  \alpha'_s ( \mu )  }
				{  \pi  }
		\frac{  \beta'_0  }
			{  \epsilon  }
	+ \left(
			\frac{  \alpha'_s ( \mu )  }
					{  \pi  }
		\right)^2
		\left(
			\frac{  \beta_0^{'2}  }
					{  \epsilon^2  }
			-
			\frac{  \beta'_1  }
					{  2 \epsilon  }
		\right)
	\;.
\end{equation}
Here $\beta'_0$ and $\beta'_1$ denote the first two coefficients of the $\beta$ function in the
light-flavours theory,
\begin{equation}
	\beta'_0 
	=  
	\frac{ 1 }{ 4 }
	\left(  11  -  \frac{ 2 }{ 3 } n_l  \right)
	\quad , \quad 
	\beta'_1 
	= 
	\frac{ 1 }{ 16 }
	\left(  102  - \frac{ 38 }{ 3 } n_l  \right)
	\;.
\end{equation}
Combining Eqs.~(\ref{eq:renormalization_full} -\ref{eq:Zalpha}) we get
\begin{equation}
	m_q^0 
	=
	m_q(\mu)
	\left\{
		   1
		- \frac{  \alpha'_s(\mu)  }{  \pi   } \frac{  1  }{   \epsilon  }
		+ \left(  \frac{  \alpha'_s(\mu)  }{  \pi   } \right)^2
			\left[
				  \frac{  45 - 2 (n_h + n_l)  }{  24 \epsilon^2  }
				+\frac{48 L_m + 10 (n_h + n_l) - 303  }{  144 \epsilon  }
			\right]
	\right\},
\end{equation}
with 
\begin{equation}
	L_m
	=
	\sum_{q=1}^{n_h}
	\log \left( \frac{  m_q(\mu)  }{  \mu  }  \right)
	\;.
\end{equation}

The next step is the renormalization of the bare strong coupling
in the effective theory, $\alpha_s^{'0}$. The relevant renormalization 
constant is given in Eq.~(\ref{eq:Zalpha}). 

We finally renormalize the bare Wilson coefficient $C_1^0( \alpha'_s, m_q )$ using
~\cite{Spiridonov:1984br,Spiridonov:1988md,Chetyrkin:1996ke}
\begin{eqnarray}
	C_1
	& = &
	\frac{  1  }
			{  1 + \alpha'_s(\mu) \frac{  \partial  }
											 {  \partial \alpha'_s(\mu)  }
			\log Z'_{\alpha}}
	C^0_1
	\nonumber \\
	& = &
	\left[
	1
	+
	\frac{  \alpha'_s(\mu)  }
			{  \pi  }
	\frac{  \beta'_0  }
			{  \epsilon  }
	+
	\left(
		\frac{  \alpha'_s(\mu)  }
			{  \pi  }
		\right)^2
	\frac{  \beta'_1  }
			{  \epsilon  }
\right]
C^0_1
	\;.
\end{eqnarray}

Our final result for the renormalized Wilson coefficient reads
\begin{eqnarray}
\label{eq:WCfinal}
C_1 
&=& 
-\frac{1}{3} \frac{ \alpha'_s(\mu) }{ \pi } 
\left\{
		n_h 
		+ \frac{11}{4} 
			 \frac{ \alpha'_s(\mu) }{ \pi } n_h 
		- \left( \frac{ \alpha'_s(\mu) }{ \pi } \right)^2 
				\left[
						-\frac{ 1877 }{ 192 } n_h 
						+ \frac  {  77  }{  576  } n_h ^2
						+  \frac{  19 L_m  }{  8  }
					\right.
				\right.
				\nonumber \\
				& & \left. \left.\phantom{QQQQQQQQQQQQQQQQQQQQQQQQ} 
						+ n_l \left(
									\frac{ 67 }{  96  } n_h
									+ \frac{ 2 L_m }{ 3 }
						\right)						
				\right]
	\right\}
	\;.
\end{eqnarray}
This is the main result of our paper.
Note that the second term in the square brackets comes 
from diagrams containing two massive quarks loops. 
For $n_h = 1$ we recover the SM Wilson coefficient 
~\cite{Chetyrkin:1997iv,Chetyrkin:1997un,Schroder:2005hy,Chetyrkin:2005ia} through order ${\cal O}({\alpha'_s}^3)$.

%Ref.~\cite{Chetyrkin:1997iv}
%
%\begin{equation}
%	\Gamma(H \to gg)
%	=
%	\frac{ G_F }{ 8 \sqrt{2} m_H } C_1^2 \, 
%	\rm{Im} \langle \left[ {\cal O}_1' \right] \left[ {\cal O}_1' \right] \rangle
%\end{equation}

At leading order in the heavy-quark expansion, and assuming a massless  bottom quark, 
the gluon fusion cross-section and the decay width of the Higgs  boson to gluons are proportional to 
the square of the Wilson coefficient.  In this limit,  their ratio with the  corresponding quantities in the Standard Model are: 
\begin{eqnarray}
	&& \frac{  \sigma(gg \to H)^{(n_h)}  }
			{  \sigma(gg \to H)^{(SM)} }
      =
	\frac{  \Gamma(H \to gg)^{(n_h)}  }
			{  \Gamma(H \to gg)^{(SM)} }
	=
	\nonumber \\ 
&& \hspace{0.5cm}
n_h^2  - \left(\frac{\alpha'_s(\mu)}{\pi}\right)^{2} n_h
	\left[
			  \frac{77}{288} n_h  (n_h-1)
			  +\left(\frac{4}{3} n_l+\frac{19}{4}   \right) \sum_q \log\left( \frac{m_q(\mu)}{m_t(\mu)} \right)
	\right] + {\cal O}(\alpha_s^{'3}).
	\nonumber \\ 
\end{eqnarray}
where $m_t$ the mass of the top-quark. 
The  ${\cal O}(\alpha_s^{'2})$ term in the above expression is  generally very small. However, 
in a realistic  phenomenological study~\cite{Anastasiou:2008tj,deFlorian:2009hc} the exact quark 
mass dependece of the cross-section as well as effects  due to 
electroweak corrections need to be accounted  for through NLO.

\section{Numerical Results for gluon fusion cross section at the Tevatron}
\label{sec:numerics}

In this Section, we present  our  numerical results for the cross-section at the Tevatron, in a Standard Model 
with four generations.  
The Wilson coefficient for the 
fourth generation model is obtained by considering three heavy quarks in Eq.~(\ref{eq:WCfinal}). 
We  set the top-quark mass  to 
\[
m_t = 170.9 \, {\rm GeV} \;.
\]
For the fourth generation we consider two scenarios, 
corresponding to fourth generation down-quark masses of: 
\[ 
m_B = 300 \,{\rm GeV},  \qquad m_B = 400 \,{\rm GeV}
\]
and  an up-quark mass given by  
\begin{equation}
\label{eq:massup}
m_T - m_B =   50 \, {\rm GeV} + 10 \log \left(\frac{m_{H}}{115 \, {\rm GeV}} \right) \, {\rm GeV}.   
\end{equation}
This choice is permitted by constraints from electroweak precision tests,  
as  described  in Ref.~\cite{Kribs:2007nz}.

The  calculation of the total cross-section differs  from the Standard Model  only in the expression of the 
Wilson coefficient for the low energy effective Lagrangian. 
We  combine Eq.~(\ref{eq:WCfinal}) with the known results for the Standard 
Model total cross-section at NNLO in the large top-mass limit of 
Refs~\cite{Harlander:2002wh,Anastasiou:2002yz,Ravindran:2003um}.

Adopting the same  approach as in Ref.~\cite{Anastasiou:2008tj},  
we first compute  the ratio of  the  NNLO and LO cross-section in the effective theory. 
We  estimate  the contribution from Feynman diagrams with only 
top-quark and  fourth generation quarks to the total cross-section, by multiplying this 
ratio  with the exact leading  order contributions of heavy quark  diagrams in the full theory. 
\begin{equation}
\sigma_{heavy}^{NNLO;(t,B,T)} \simeq \left(  \frac{
\sigma^{NNLO;(t,B,T)}
}{
\sigma^{LO;(t,B,T)}
} 
\right)_{effective} \sigma_{exact}^{LO;(t,B,T)}
\end{equation}
These contributions are enhanced by roughly a factor  of 9 with respect to the 
corresponding Standard Model results, since 
\begin{equation}
\sigma_{exact}^{LO;(t,B,T)} \simeq 9 \; \sigma_{exact}^{LO;(t)} 
\end{equation}
within a few  percent. 

Contributions from diagrams  with bottom quark loops are small and  we compute them exactly 
through the  NLO order approximation~\cite{Anastasiou:2009kn,Spira:1995rr}.     
In comparison to their Standard Model counterparts, the most important interference terms of diagrams with bottom quarks only 
 and  diagrams with any of the  heavier quarks are  only enhanced by roughly a  factor of  three. 
Therefore, these  contributions  are  suppressed  by roughly a  factor  of $\sim 3/9$ in this model. 

Finally, we include two-loop electroweak corrections from light-quark loops from the first  two generations~\cite{Aglietti:2004nj} in the complex mass scheme, 
and the corresponding three-loop mixed QCD and electroweak corrections as in Ref.~\cite{Anastasiou:2008tj}.   These are enhanced by a factor  of roughly 3 in comparisoin to 
the Standard Model, and  are therefore  suppressed by a factor of  $1/3$ in this model. We ignore  electroweak corrections with quarks  from the third and fourth generation, which are  
already found  to be very small in the Standard Model~\cite{Actis:2008ug} for the Higgs mass range  accessible at the Tevatron, when a  complex mass scheme  is employed. 

In Table~\ref{tab:results}, we present the  cross-section at a renormalization and factorization scale $\mu = \mu_f=\mu_r = m_{H}/2$ and 
estimate the scale variation error by varying the common scale $\mu$ in the interval $\left[\frac{m_H}{4} , m_{H}\right]$.  
We  use  the MSTW2008 NNLO parton distribution functions~\cite{Martin:2009iq}, and  compute the uncertainty 
(with $90\% \, {\rm CL}$) to the cross-section due to  the parton distribution functions (including the parametric uncertainty of the  
value of $\alpha_s$) according to Ref.~\cite{Martin:2009bu}.
\begin{table}[h]
\begin{center}
\begin{tabular}{|c|c|c|c|c|}
\hline
$m_H$(GeV)&$\sigma_{(1)}(fb)$&$\sigma_{(2)}(fb)$&$\frac{\delta \sigma}{\sigma} ({\rm pdf}+\alpha_s) \%$&$\frac{\delta \sigma}{\sigma}({\rm scale})\%$\\
\hline
\hline
110& 12384& 12308& ${+12\%}\, , \,{-11\%}$& ${+12\%}\, , \,{-8\%}$\\ \hline
115& 10798& 10725& ${+12\%}\, , \,{-11\%}$& ${+12\%}\, , \,{-8\%}$\\ \hline
120& 9449.9& 9384.3& ${+12\%}\, , \,{-11\%}$& ${+12\%}\, , \,{-8\%}$\\ \hline
125& 8298.8& 8240.0& ${+12\%}\, , \,{-12\%}$& ${+12\%}\, , \,{-8\%}$\\ \hline
130& 7314.0& 7258.7& ${+12\%}\, , \,{-12\%}$& ${+12\%}\, , \,{-8\%}$\\ \hline
135& 6465.1& 6414.2& ${+12\%}\, , \,{-12\%}$& ${+12\%}\, , \,{-8\%}$\\ \hline
140& 5731.4& 5684.1& ${+13\%}\, , \,{-12\%}$& ${+12\%}\, , \,{-8\%}$\\ \hline
145& 5094.6& 5050.4& ${+13\%}\, , \,{-12\%}$& ${+12\%}\, , \,{-8\%}$\\ \hline
150& 4540.5& 4498.5& ${+13\%}\, , \,{-12\%}$& ${+12\%}\, , \,{-8\%}$\\ \hline
155& 4055.6& 4017.6& ${+13\%}\, , \,{-12\%}$& ${+12\%}\, , \,{-8\%}$\\ \hline
160& 3630.2& 3595.1& ${+13\%}\, , \,{-13\%}$& ${+12\%}\, , \,{-8\%}$\\ \hline
165& 3253.7& 3220.7& ${+14\%}\, , \,{-13\%}$& ${+12\%}\, , \,{-8\%}$\\ \hline
170& 2924.1& 2893.2& ${+14\%}\, , \,{-13\%}$& ${+12\%}\, , \,{-8\%}$\\ \hline
175& 2633.9& 2604.4& ${+14\%}\, , \,{-13\%}$& ${+12\%}\, , \,{-8\%}$\\ \hline
180& 2376.7& 2348.9& ${+14\%}\, , \,{-13\%}$& ${+12\%}\, , \,{-8\%}$\\ \hline
185& 2147.2& 2121.5& ${+15\%}\, , \,{-13\%}$& ${+12\%}\, , \,{-8\%}$\\ \hline
190& 1943.9& 1919.7& ${+15\%}\, , \,{-14\%}$& ${+12\%}\, , \,{-8\%}$\\ \hline
195& 1763.2& 1740.2& ${+15\%}\, , \,{-14\%}$& ${+12\%}\, , \,{-8\%}$\\ \hline
200& 1601.8& 1580.0& ${+15\%}\, , \,{-14\%}$& ${+12\%}\, , \,{-8\%}$\\ \hline
205& 1457.5& 1436.7& ${+16\%}\, , \,{-14\%}$& ${+12\%}\, , \,{-8\%}$\\ \hline
210& 1328.1& 1308.4& ${+16\%}\, , \,{-14\%}$& ${+12\%}\, , \,{-8\%}$\\ \hline
215& 1212.0& 1193.2& ${+16\%}\, , \,{-14\%}$& ${+12\%}\, , \,{-8\%}$\\ \hline
220& 1107.7& 1089.6& ${+16\%}\, , \,{-15\%}$& ${+12\%}\, , \,{-8\%}$\\ \hline
225& 1013.6& 996.33& ${+17\%}\, , \,{-15\%}$& ${+12\%}\, , \,{-8\%}$\\ \hline
230& 928.61& 912.21& ${+17\%}\, , \,{-15\%}$& ${+12\%}\, , \,{-8\%}$\\ \hline
235& 852.00& 836.33& ${+17\%}\, , \,{-15\%}$& ${+12\%}\, , \,{-8\%}$\\ \hline
240& 782.52& 767.44& ${+17\%}\, , \,{-15\%}$& ${+12\%}\, , \,{-8\%}$\\ \hline
245& 719.64& 705.19& ${+18\%}\, , \,{-15\%}$& ${+12\%}\, , \,{-8\%}$\\ \hline
250& 662.60& 648.81& ${+18\%}\, , \,{-16\%}$& ${+12\%}\, , \,{-8\%}$\\ \hline
255& 610.74& 597.51& ${+18\%}\, , \,{-16\%}$& ${+12\%}\, , \,{-8\%}$\\ \hline
260& 563.53& 550.90& ${+19\%}\, , \,{-16\%}$& ${+12\%}\, , \,{-8\%}$\\ \hline
265& 520.60& 508.52& ${+19\%}\, , \,{-16\%}$& ${+12\%}\, , \,{-8\%}$\\ \hline
270& 481.49& 469.93& ${+19\%}\, , \,{-16\%}$& ${+12\%}\, , \,{-8\%}$\\ \hline
275& 445.86& 434.72& ${+20\%}\, , \,{-16\%}$& ${+12\%}\, , \,{-8\%}$\\ \hline
280& 413.24& 402.68& ${+20\%}\, , \,{-17\%}$& ${+12\%}\, , \,{-8\%}$\\ \hline
285& 383.56& 373.28& ${+20\%}\, , \,{-17\%}$& ${+12\%}\, , \,{-8\%}$\\ \hline
290& 356.39& 346.53& ${+21\%}\, , \,{-17\%}$& ${+12\%}\, , \,{-8\%}$\\ \hline
295& 331.53& 322.04& ${+21\%}\, , \,{-17\%}$& ${+12\%}\, , \,{-8\%}$\\ \hline
300& 308.70& 299.71& ${+21\%}\, , \,{-17\%}$& ${+12\%}\, , \,{-8\%}$\\ \hline
\end{tabular}

\end{center}
\caption{ 
\label{tab:results}
The NNLO cross-section for  Higgs production via gluon fusion at the TEVATRON. $\sigma_{(1)}$ corresponds  to $m_{B} = 300 {\rm GeV}$ and $\sigma_{(2)}$ 
to    $m_{B} = 400 {\rm GeV}$. The mass  of the fourth generation up quark is given by Eq.~(\ref{eq:massup})
}
\end{table}

The scale variation uncertainty and the uncertainty from the parton distributions are the dominant uncertainties. Essentially,  they are 
the same as in the Standard Model cross-section with only three generations. We  note that  Ref.~\cite{Anastasiou:2008tj} preceded 
the  release  of  Ref.~\cite{Martin:2009bu} where it  became  possible  to include the  uncertainty of the $\alpha_s$ value  in the fitted 
parton densities and the corresponding parton density uncertainty was  estimated  to be smaller.

\section{Conclusions}
\label{sec:conclusions}

In this paper, we constructed an effective field  theory for  extensions  of the Standard Model with 
many heavy quarks coupling to the Higgs  boson via top-like Yukawa interactions.  
We  have  computed the required  Wilson coefficient of the $-\frac{H}{4v} G_{\mu \nu} G^{\mu \nu}$ operator 
through NNLO in the strong coupling expansion.  
We found the  result
\begin{eqnarray}
\label{eq:WCfinalfinal}
C_1 
&=& 
-\frac{1}{3} \frac{ \alpha'_s(\mu) }{ \pi } 
\left\{
		n_h 
		+ \frac{11}{4} 
			 \frac{ \alpha'_s(\mu) }{ \pi } n_h 
		- \left( \frac{ \alpha'_s(\mu) }{ \pi } \right)^2 
				\left[
						-\frac{ 1877 }{ 192 } n_h 
						+ \frac  {  77  }{  576  } n_h ^2
						+  \frac{  19 L_m  }{  8  }
					\right.
				\right.
				\nonumber \\
				& & \left. \left.\phantom{QQQQQQQQQQQQQQQQQQQQQQQQ} 
						+ n_l \left(
									\frac{ 67 }{  96  } n_h
									+ \frac{ 2 L_m }{ 3 }
						\right)						
				\right]
	\right\}
	\;.
\end{eqnarray}
This  result can be utilized by the ongoing experimental studies at the Tevatron and the LHC to 
constrain such models.  
The cross-section and the decay width are enhanced by roughly the square of the number of  heavy quarks   
%\begin{equation}
%\sigma \sim n_h^2 \sigma_{SM}, 
%\end{equation}
with respect to the Standard Model.  
The Tevatron has put  stringent  limits on the Standard Model Higgs  boson gluon fusion 
cross-section~\cite{Aaltonen:2010cm}.  Equivalent studies can be performed  in  models with additional quarks. 
We  have presented numerical results  for the gluon fusion cross-section at the Tevatron 
in a non-minimal Standard Model with four generations. The theoretical uncertainties of the cross-section are 
practically independent of the number of heavy quarks and very similar to the Standard Model.

\section*{Acknowledgments}
We thank Achilleas Lazopoulos for many useful discussions  and his help. We thank Giuliano Panico for 
comments on the manuscript. This research is supported by the Swiss National Science Foundation under 
contracts 200020-116756/2  and  200020-126632
%%%%%%%%%%%%%%%%%%%%%%%%%%%%%%%%%%%%%%%%%%%%%%%%%
%%%%% bibliography
%%%%%%%%%%%%%%%%%%%%%%%%%%%%%%%%%%%%%%%%%%%%%%%%%%

\end{document}